\documentclass[10pt,conference,twocolumn]{IEEEtran}
\usepackage{amsthm}

\theoremstyle{definition}

\theoremstyle{definition}

\usepackage{amssymb,amscd}
\usepackage{graphicx}
\usepackage{braket}
\usepackage{dsfont}
\usepackage{color}
\usepackage{tikz}
\usetikzlibrary{arrows,positioning}
\usetikzlibrary{shapes.geometric}
\usetikzlibrary{patterns}
\usepackage{mathrsfs}
\usepackage{mathtools}
\usepackage{tcolorbox}
\usepackage[utf8]{inputenc}
\usepackage{enumerate}
\usepackage[english]{babel}
\usepackage{lipsum} 
\usepackage{cite}
\usepackage{amsmath,amssymb,amsfonts}
\usepackage{textcomp}

\newcommand\kket[1]{\vert#1\rangle\!\rangle}
\newcommand\bbra[1]{\langle\!\langle#1\vert}

\newcommand{\myi}{\ensuremath{\mathrm{i}}}

\usepackage{nicefrac}
\usepackage{physics}
\usepackage{bm}

\numberwithin{equation}{section}

\DeclareMathOperator{\Diag}{diag}
\DeclareMathOperator{\Span}{span}

\DeclareMathOperator{\Ring}{\mathbb{Z}}
\DeclareMathOperator{\Fidelity}{\mathbf{F}}

\DeclareMathOperator{\closure}{\mathrm{closure}}
\newcommand{\spanbraket}[1]{\langle#1\rangle}
\newcommand{\eigenvalue}{\eta}
\newcommand{\repx}{X}
\newcommand{\isomorphic}{\cong}
\newcommand{\character}{\chi}

\newcommand{\permutation}{\sigma}

\newcommand{\qudim}{d}
\newcommand{\rootunit}{\omega}

\newcommand{\irrepindex}{\varpi}
\newcommand{\channel}{\ensuremath{\mathcal{E}}}

\newcommand{\hilbertspace}{\mathscr{H}}

\newcommand{\finitegroup}{\mathcal{G}}
\newcommand{\finitesubgroup}{\mathcal{H}}
\newcommand{\integers}{\mathbb{Z}}
\newcommand{\setofmatrices}{\mathscr{M}}

\newcommand{\densitymatrix}{\rho}
\newcommand{\ringelement}{r}
\newcommand{\krausop}{A}
\newcommand{\setofkraus}{\mathscr{A}}
\newcommand{\twirlop}{\mathcal{T}}
\newcommand{\circuitdepth}{m}
\newcommand{\identity}{\mathbb{I}}

\newcommand{\unitarygroup}{\mathrm{U}}
\newcommand{\unitarymatrix}{U}
\newcommand{\setunitary}{\mathcal{U}}
\newcommand{\chimatrix}{\chi}
\newcommand{\chientry}{\chi_{00}}
\newcommand{\groupelement}{g}
\newcommand{\anothergroupelement}{h}

\newcommand{\tmatrix}{T}
\newcommand{\specialsub}{\mathcal{C}}
\newcommand{\pauligroup}{\mathcal{P}}

\newcommand{\spampara}{a}
\newcommand{\spamparb}{b}
\newcommand{\powerT}{\mathrm{p}}
\newcommand{\singlematrix}{\delta}
\newcommand{\symmetricgroup}{S}
\newcommand{\normaliser}{\mathcal{N}}
\DeclareMathOperator*{\composite}{\bigcirc}
\usepackage{xspace}

\newcommand{\noiseT}{\channel_\tmatrix}
\newcommand{\noiseC}{\channel_\specialsub}
\usepackage{graphicx}

\usepackage{tikz}
\usetikzlibrary{quantikz2}
\usetikzlibrary{external}
\usepackage{import}
\usepackage{xifthen}
\usepackage{pdfpages}
\usepackage{transparent}

\IEEEoverridecommandlockouts
\begin{document}

\title{Qudit non-Clifford interleaved benchmarking\\
\thanks{NSERC and Government of Alberta}
}

\author{\IEEEauthorblockN{David Amaro-Alcal\'a}
\IEEEauthorblockA{\textit{Institute for Quantum Science and Technology} \\
\textit{University of Calgary}\\
Calgary, Canada \\
david.amaroalcala@ucalgary.ca}
\and
\IEEEauthorblockN{Barry C. Sanders}
\IEEEauthorblockA{\textit{Institute for Quantum Science and Technology} \\
\textit{University of Calgary}\\
Calgary, Canada \\
sandersb@ucalgary.ca}
\and
\IEEEauthorblockN{Hubert de Guise}
\IEEEauthorblockA{\textit{Department of Physics} \\
\textit{Lakehead University}\\
Thunder Bay, Canada \\
hdeguise@lakeheadu.ca}
}

\maketitle


\begin{abstract}
We introduce a scheme to characterise a qudit T gate that has different noise than a set of Clifford gates.
We developed our scheme through representation theory and ring theory to generalise
non-Clifford interleaved benchmarking to qudit systems.
By restricting to the qubit case, we recover the dihedral benchmarking
scheme.
Our characterisation scheme provides experimental physicists a practical method for
characterising universal qudit gate sets
and
advances randomised benchmarking research by
providing 
the characterisation
of a complete qudit library.
\end{abstract}


\section{Introduction}
Scalable qudit quantum computing requires 
low gate-error rates, as dictated by fault-tolerant
schemes~\cite{Aharonov_Ben-Or_2008}.
As assessing the diamond norm is infeasible, experimental characterisation
relies instead on the
average gate fidelity.
A method to estimate the fidelity, activity also called characterisation, of qudit gates is necessary.
Whereas the individual characterisation of a qubit gate is
known~\cite{MagesanEaswar2012Emoq,Harper_Flammia_2017},
the
analogous technique for qudits is missing.
In our work, we introduce a method to estimate the fidelity of non-Clifford
qudit gates for gates with different noise than Clifford and Pauli gates;
we refer to this procedure as individual
characterisation~\cite{MagesanEaswar2012Emoq}.

Our research is motivated by scalable qudit quantum computing. 
We develop a scheme to determine if universal qudit gates
are fault-tolerant.
We present our non-Clifford interleaved benchmarking extension to characterise
individual universal qudit gates.

Currently, there are two main approaches for individual gate characterisation.
Gate-set tomography stands out as the most comprehensive
yet resource-expensive approach for this
task~\cite{merkel2013}.
A popular alternative to tomography is interleaved benchmarking
(IB)~\cite{Elben2023,MagesanEaswar2012Emoq}.
IB scheme relies on a pre-characterised library, which we call reference library~\cite{MagesanEaswar2012Emoq}.
The reference library is used to characterise individual gates, regardless of whether
these individual gates are members of the library or not.

Qubit IB itself has some variants.
One variant employs the whole Clifford gate set as an auxiliary
library~\cite{Harper_Flammia_2017,Garion2021}.
Another uses a smaller subset of the Clifford gate set, at the cost of
requiring two randomised benchmarking experiments;
this is called non-Clifford IB~\cite{dugas2015,Onorati19}.

Our focus is on extending non-Clifford IB.
We employ a subset of our own library~\cite{amaro2024a,amaro2024b} for the characterisation of a gate set
including a non-Clifford gate. 
Our method requires fewer gates than previous methods employing the whole
Clifford library.
Moreover, the circuit design complexity remains the same as the 
circuit used in standard randomised benchmarking schemes~\cite{Magesan2012}.

Individual characterisation of qudit non-Clifford gates using randomised benchmarking
schemes is required.
However, there is no generalisation for systems beyond qubits; even for qubits,
the non-Clifford interleaved benchmarking (IB) scheme has not been formalised.
In our work, we present a formalisation of the non-Clifford interleaved benchmarking scheme.
Based on our formalisation, we extend the qubit-level non-Clifford IB scheme to
characterise universal qudit gates.

\section{Background}\label{sec:background}
In this section, we review concepts from mathematics and quantum information.
From mathematics, we revisit tools from ring theory and representation theory, such as cosets and representations.
We use Howell's algorithm for row-reducing matrices with entries in an arbitrary ring.
In the realm of quantum information, we review the definitions of quantum channels and states, along with their representations.
We also recall the notions of average and sequence fidelity.

\subsection{Qudits and multi-valued logic}
Bits are the basic unit of computation in binary logic.
The original model of quantum computing was based on a quantum analogy to
bits---the
qubit~\cite{Benioff1982a,Benioff1982b,Zurek1984,Peres1985,Feynman1986,Landauer1986-qi,Margolus1986-co}.

Nowadays, in the same way there is multi-valued logics~\cite{choe2023},
quantum information processors using multi-level systems,
which are known as qudits,
are being
constructed~\cite{Ringbauer2022,Lindon2022,Morvan_Ramasesh_Blok_Kreikebaum_OBrien_Chen_Mitchell_Naik_Santiago_Siddiqi_2021}.
The description of these multi-valued quantum units of computation is presented
in the following subsection.

\subsection{Quantum}
In this section we recall key concepts in quantum information theory.
We begin with states, measurements, and channels.
The vectorisation, or reshaping, is explained.
We also explain how channels emerge from unitary operations.
Additionally, we discuss gate universality, highlighted by examples relevant for
our IB scheme. 


We work with a \(\qudim\)-dimensional Hilbert space \(\hilbertspace_\qudim \coloneqq
\Span(\ket{0},\ldots,\ket{\qudim-1})
\isomorphic \mathbb{C}^{\qudim}
\).
Let \(\setofmatrices_\qudim\) denote the set of invertible \(\qudim\times \qudim\) complex
matrices.
A density matrix \(\densitymatrix \in\setofmatrices_\qudim\) is an hermitian matrix
with trace one~\cite{MikeAndIke,HeinosaariTeiko2012Tmlo}.

Given \(\krausop\in\setofmatrices_\qudim\),
\(\krausop^{\dagger}\) denotes the hermitian conjugate and \(\overline{\krausop}\) 
denotes \(\krausop\) complex conjugate.

A channel \(\channel\colon \setofmatrices_\qudim\to\setofmatrices_\qudim\) is a completely positive trace preserving
mapping (CPTP)~\cite{MikeAndIke,kraus1983}.
Given a finite subset \(\setofkraus \subsetneq \setofmatrices_\qudim\),
we can define a mapping denoted \(\channel_{\setofkraus}\) as
\begin{equation}
\channel_{\setofkraus}(\densitymatrix)
=
\sum_{\krausop\in\setofkraus} \krausop \densitymatrix \krausop^{\dagger}.
\end{equation}
A mapping \(\channel_{\setofkraus}\) is CPTP~\cite{kraus1983} iff
\begin{equation}
\sum_{\krausop\in\setofkraus} \krausop^{\dagger}\krausop = \identity_\qudim.
\end{equation}
We now recall a convenient way to deal with channels and states.

We introduce the vectorisation of a density matrix and a
channel~\cite{HeinosaariTeiko2012Tmlo, karolgeometry2017}.
The vectorisation operation reshapes \(\densitymatrix\in\setofmatrices\) to a \(\qudim^2\)-dimensional vector.
The vectorisation of 
the action of a mapping \(\channel\)
acting on
\(\densitymatrix\) is defined as
\begin{equation}\label{eq:vectorisation}
\kket{\channel(\densitymatrix)}
=
\sum_{\krausop\in\setofkraus}
\krausop\otimes\overline{\krausop} \kket{\densitymatrix}.
\end{equation}
We now discuss the notion of universality of quantum gates.

There is an important class of mappings inside \(\setofmatrices_\qudim\): universal mappings.
Universality is a notion defined for a discrete set of unitary
matrices~\cite{MikeAndIke}.
A set of \(\qudim\times \qudim\) unitary matrices \(\setunitary\) is universal if 
\(\closure\spanbraket{\setunitary} = \unitarygroup(\qudim)\).
The notion of universality is then ``lifted'' to gates using the channel
associated with a unitary matrix.

Universality is also relevant in the realm of quantum information.
An arbitrary gate can be approximated using a finite combination of elements in
the universal set.
Importantly, as a consequence of Solovay-Kitaev theorem and its qudit
generalisations~\cite{Aharonov_Ben-Or_2008,Kitaev_1997,Dawson_Nielsen_2005}, 
universality
has
two additional advantages: 
the circuit depth and the running time for the circuit design are
efficient~\cite{Dawson_Nielsen_2005}.

For each square \(\qudim\)-dimensional unitary matrix \(\unitarymatrix\), we can
define a CPTP mapping~\cite{MikeAndIke}.
Taking 
\(\setofkraus = \{\unitarymatrix\}\),
\(\channel_{\setofkraus}(\rho) \coloneqq 
\unitarymatrix\rho\unitarymatrix^{\dagger}\), where \(\rho\)
is a density matrix.
Thus, given a universal set of matrices \(\setunitary\),
we call the set of channels \(
\{\channel_{\unitarymatrix} \colon \unitarymatrix\in \setunitary\}\) 
universal.

There is an important gate set with origins in quantum
teleportation~\cite{calderbank1998,Gottesman1999}---the Clifford gate set.
This gate set is generated by the quantum Fourier transform gate and a phase
gate~\cite{Tolar_2018}.
Standard randomised benchmarking characterises
them~\cite{Jafarzadeh_Wu_Sanders_Sanders_2020},
therefore in this work we assume
any Clifford gate has been characterised

Two gates suffice to attain single-qudit universality: the discrete Fourier
transform matrix H (Chrestenson) and a T qudit
generalisation~\cite{chrestenson1955}.
H is a Clifford gate and T is a diagonal member of the third level of the
Clifford hierarchy.
Together, H and T generate a dense subgroup of \(\unitarygroup(\qudim)\).

\subsection{Randomised benchmarking}
A gate is characterised if we know its average gate fidelity.
Randomised benchmarking is the preferred method for gate characterisation.
This preference is due to its low circuit design cost and
simple data analysis~\cite{Elben2023}.

A randomised benchmarking scheme has three stages:
design, experimental or simulation, and data analysis.
The design stage corresponds to the part where given a set of gates to characterise, a
set of expressions for the sequence and gate fidelity are presented. 
The experimental stage is where the data is collected.
The data analysis is where the data and the output of the design stage are used
to determine the quality of a set of gates.

The design stage 
consists of obtaining expressions and a set up from experimental
resources; these resources are mainly gates, state preparation, and
measurements.
The set of formulae are the expressions for the gate and sequence fidelity.
The formulae are then used in the data-analysis stage, along the experimental
results, to characterise the gates.

The experimental stage uses the experimental scheme to estimate a binary array:
0 for no detection and 1 for detection.
Each entry in the binary array
corresponds to the output of the process: preparation, application of gates, and measurement
sequence.
This binary array, and the formulae obtained in the design part, are used to
characterise the gate set.

The data-analysis part concludes the execution of a RB scheme.
It takes as input the experimental data and the formulae of the design part.
The outcome is an estimate of the gate fidelity for the gate set or gate-set
member.
This is the characterisation of a gate set.

Our work deals with the design part of RB schemes.
We provide formulae for the sequence fidelity and gate fidelity, an experimental
procedure, and a fitting procedure.
These three outcomes form our interleaved benchmarking scheme.

The average gate fidelity quantifies the quality of quantum gates.
For a given \(\channel_\setofkraus\), we have the following simple
formula~\cite{NIELSEN2002249}: 
\begin{equation}\label{eq:simple-formula}
\Fidelity(\channel)
=
\frac{\qudim\tr(\sum_{\krausop\in\setofkraus} \krausop\otimes\bar{\krausop}) +
\qudim^2}{\qudim^2(\qudim+1)}.
\end{equation}
From Eq.~\eqref{eq:simple-formula} we learn that to estimate the fidelity, only
the trace of a channel is necessary.
In \S\ref{sub:group-rep-theory} we recall another concept used in our RB scheme:
sequence fidelity.

\subsection{Sequence fidelity}\label{sub:group-rep-theory}
Representation theory plays a fundamental role in randomised benchmarking.
To fix notation, we recall some group theory concepts.
Given a group \(\finitegroup\) and a subgroup
\(\finitesubgroup\subset\finitegroup\), we define a coset.
A coset is defined in terms of a  \(\groupelement\in\finitegroup\)
and  \(\finitesubgroup\) as  \(\groupelement\finitesubgroup \coloneqq
\{\groupelement\anothergroupelement\colon \anothergroupelement\in\finitesubgroup\}\).
We denote the conjugate of~\(\groupelement\) by \(\anothergroupelement\)
as  \(\groupelement^{\anothergroupelement} \coloneqq
\anothergroupelement\groupelement \anothergroupelement^{-1}\).
Similarly, for a  finite subgroup \(\finitesubgroup\),
\(\finitesubgroup^{\groupelement} \coloneqq
\{\anothergroupelement^{\groupelement}\colon
\anothergroupelement\in\finitesubgroup\}\).

We introduce an important representation for RB schemes.
Consider a unirrep implicitly defined by a set of \(\qudim\)-dimensional unitary matrices
\(\mathcal{U}\).
We call the representation defined as \(u\mapsto u\otimes\bar{u}\colon u\in \mathcal{U}\)
the channel associated with a unitary matrix~\(u\in \mathcal{U}\).

Labelling channels by group elements (or their representation) is key part of RB schemes.
From now on, let \(\groupelement\) denote both a group element and its
\(\qudim\times\qudim\) unitary representation.
Assume for each \(\groupelement\in\finitegroup\) there is a channel
\(\channel_\groupelement\).
Henceforth, for gates labelled by group elements, we refer to the
composition and inverse operation as the group operation.
Whereas a channel \(\channel\) may not necessary have an inverse,
we can use the label \(g\) to define an alternative inverse operator;
the inverse of~\(\channel_\groupelement\) is the channel labelled by the inverse
of \(g\):
\(\channel_{\groupelement^{-1}}\).

We now define the inversion gate for a sequence of gates.
Consider a set of group elements \(\{g_i\}\), a multiset of \(\circuitdepth\) gates \(\channel_{\groupelement_i}\),
a state \(\rho\), and a measurement \(E\).
By labelling gates with group elements \(\groupelement_i\), we can define the inversion gate
associated with the inverse of the sequence  \(\composite_i \channel_{\groupelement_i}\) as~\(\channel_{(\composite_i \groupelement_i)^{-1}}\).
Using the inversion gate we define the sequence fidelity for a sequence of
gates.
The version of the sequence fidelity we use is
\begin{equation}
  \Pr(m;\rho, E) \coloneqq 
  \bbra{E}
\channel_{(\composite_i \groupelement_i)^{-1}}
\composite_i \channel_{\groupelement_i}
  \kket{\rho}.
\end{equation}
At this point we need to briefly digress
to introduce notions of ring theory necessary in our
scheme.

\subsection{Ring theory}

To fix notation, we recall several facts on ring theory~\cite{dummit2018}
We rely on rings such as \(\Ring_k\) where  \(k\in \integers\).

We are interested on row-reduced form of a matrix with entries in certain ring
\(\Ring_k\).
The non-trivial part of such procedure is that, in distinction with a field, not every
element 
has a multiplicative inverse in \(\Ring\).

Fortunately, there is an algorithm, known as Howell's algorithm, that carries out
this row-reduction procedure.

This algorithm transforms a matrix with entries in some ring~\(\Ring_i\) and
produces a matrix resembling a row-reduced form.
This procedure is important for two tasks relevant in our scheme:
to determine a basis from a given set of
vectors and determine if a vector is 
a linear combination of some other list of vectors.

\subsection{Fidelity estimate by interleaved benchmarking}

In this section,
we introduce the ingredients of our 
qudit generalisation of non-Clifford interleaved benchmarking.
We focus on qutrit and ququart cases, as these are of current
interest~\cite{Blok_Ramasesh_Schuster_OBrien_Kreikebaum_Dahlen_Morvan_Yoshida_Yao_Siddiqi_2021,Lindon2022}.
We comment on the
\(\chimatrix\)-representation~\cite{HeinosaariTeiko2012Tmlo,chuang1997} for an expression of the AGF,
and then discuss,
in general, interleaved benchmarking for non-Clifford gates.


The \(\chimatrix\)-representation~\cite{MikeAndIke} of a quantum channel is used to describe an
approximation~\cite{Kimmel2014}.
We can approximate the
\(\chientry\) of 
the composition of two channels
and the product of the \(\chientry\) of each channel.
The approximation is
\begin{equation}\label{eq:approximation-chi-00}
  \chientry(\channel\channel') \approx \chientry(\channel) \chientry(\channel').
\end{equation}

From \(\chientry(\channel)\) the average gate fidelity can be obtained.
For a channel \(\channel\) in the \(\character\)-representation, the AGF is
\begin{equation}\label{eq:relation-chi-average-gate-fidelity}
\Fidelity(\channel) = 
\frac{d\chientry(\channel)}{d+1}
+
\frac{1}{d+1},
\end{equation}
where $\chientry(\channel)$ is a coefficient of the 
$\chimatrix$-representation of a channel~\cite{HeinosaariTeiko2012Tmlo}.

\section{Method}
This section's objective is to compute a sequence of gates with effective noise
corresponding to the composition of the noise of a T gate and a set of already
characterised gates.
We use the tools introduced in \S\ref{sec:background}.
First, we construct a Clifford-like gate set. Then we construct a circuit
interleaving T and members of the Clifford-like gate set.
The sequence fidelity obtained from sampling gates according to our circuit
design is then shown to be that of sampling from the Clifford-like gate set with
composite noise.
The noise corresponds to the composition of the noise of T and Clifford-like
gates.

\subsection{Formalisation of interleaved benchmarking}
In this subsection we introduce our generalisation of interleaved benchmarking
for non-Clifford gates.
This subsection is divided in two parts. The first is about the general
conditions of the group represented by a set of gates. 
The key part is using gates labelled by coset representatives.
Then we introduce our generalisation relying in coset representatives.

Consider a group \(\finitegroup\) and two subgroups  \(\finitesubgroup,
\finitesubgroup'\).
Further assume each element \(\groupelement\in\finitegroup\) is written in terms of two
elements in \(\anothergroupelement\in\finitesubgroup\)
\(\groupelement'\in\finitesubgroup'\): \(g = h g'\).
This decomposition may not be unique; for our purposes, it is only necessary
that the decomposition exists.

The central idea of IB~\cite{MagesanEaswar2012Emoq,dugas2015} is to obtain the
average gate fidelity of \(\channel_{\finitegroup}\)
 and then use an
 approximation
\(
\Fidelity(\channel_{\finitegroup})
=
\Fidelity(\channel_{\finitesubgroup})
\Fidelity(\channel_{\finitesubgroup'})
+\mathrm{error}(\Fidelity(\channel_{\finitesubgroup}), \Fidelity(\channel_{\finitesubgroup'}))
\)~\cite{Kimmel2014}.

We assume the fidelity of the gates in \(\finitesubgroup'\) is known.
We want to characterise a single gate in  \(\finitesubgroup\).


We now illustrate the application of the approximation in Eq.~\eqref{eq:approximation-chi-00} within the IB context.
Suppose \(U\) has been characterised, that is,  \(\Fidelity(\channel_U)\) is known.
Further assume we characterise some elements of \(\finitegroup\) in such a way that we estimate~\(\Fidelity(\channel_U \channel_V)\).
Then, by using the approximation Eq.~\eqref{eq:approximation-chi-00}, we compute
\(\Fidelity(\channel_V)\) and thus characterise \(V\).



\subsection{Construction of the Clifford-like subgroup}

In this subsection, we construct a group \(\specialsub\); 
\(\specialsub\) is formed by Clifford and cyclic mappings.
We also introduce a representation of \(\specialsub\) that induces a
bi-parametric twirl~\cite{MagesanEaswar2012Emoq}.
Given the availability of Clifford RB schemes, 
it is sensible to consider~\(\specialsub\)  already benchmarked.

For concreteness, we write down the \(T\) matrices we use in our
work~\cite{Watson2015}.
The \(T\) matrix we use for qutrits is
\begin{equation}
T = \Diag(1,\rootunit_9,\rootunit_9^{8}),
\end{equation}
where \(\rootunit_\qudim \coloneqq \exp(2\pi \myi/\qudim)\).
For ququarts
\begin{equation}
T = \Diag(1,\rootunit_8^{5},1,\rootunit_8^{7}).
\end{equation}
The particular \(\tmatrix\) used is clear from the context so we do not require
additional notation.

We introduce an auxiliary representation of the symmetric group \(S_\qudim\).
First, let \(\singlematrix_{i,j}\) denote the \(\qudim\)-dimensional matrix with only the
entry \(i,j\) equal to one and the rest zero.
Let \(\permutation\in \symmetricgroup_\qudim\) denote a permutation.
We define the representation of  \(\permutation\) as
\begin{equation}\label{eq:repx}
\repx_\permutation \coloneqq \sum_i \singlematrix_{\permutation(i),i}.
\end{equation}


We start with the preliminaries of the construction of the representation of \(\specialsub\).
First, compute all the permutations of
the diagonal entries of \(\tmatrix\); denote this set by \(\mathcal{T}\).
Let \(\finitegroup\) be the group generated by \(\mathcal{T}\).
Let \(\pauligroup_\qudim\) denote the qudit Pauli group~\cite{Tolar_2018}.
Compute the normaliser of the Pauli group by \(\mathcal{T}\).
Let the set of generators of the normaliser be \(\mathcal{N}\).

We are ready to construct the representation of the Clifford-like gates.
First, compute \(\specialsub' \coloneqq \spanbraket{ X_\sigma T
X_\sigma^{\dagger}\colon \sigma\in S_\qudim}\).
Thus \(\specialsub'\) is the group generated by the set of matrices
corresponding to permutation of the diagonal entries of \(\tmatrix\).
Then compute the normaliser of the Pauli group by \(\specialsub'\):
\(\normaliser \coloneqq \{\groupelement\in \specialsub'\colon
\pauligroup_\qudim^{\groupelement} = \pauligroup_\qudim\}\)---a subset of the
Clifford group.

The last step in the computation of the representation is to choose the
generators of \(\normaliser\) and add, to \(\specialsub\), the image of the representation \(\repx\)
in Eq.~\eqref{eq:repx}.
Therefore, our Clifford-like representation is 
defined as \(\specialsub \coloneqq
\spanbraket{ \normaliser, X, X_{01}}\),
where \(X \coloneqq  X_{(1\cdots \qudim)}\) and \(X_{01} \coloneqq X_{(12)}\).

We now are explicit and present \(\specialsub\) for qutrit and ququart systems.
The elements of \(\specialsub\) for qutrit systems are
\begin{equation}
X_\sigma \tmatrix_0^{\ringelement_0} \tmatrix_1^{\ringelement_1}
\end{equation}
where \(\tmatrix_0 = \Diag(\rootunit_9^{2}, \rootunit_9^{5}, \rootunit_9^{2})\),
\(\tmatrix_1 = \rootunit_3\identity\),
and \(\ringelement_0,\ringelement_1\in\Ring_9\).
On the other hand, the generators of the ququart part are:
\begin{subequations}
\begin{align}\label{eq:generators-diagonal-ququart}
  \tmatrix_0 &=\Diag(1,\rootunit_4,\rootunit_4,1),\\
\tmatrix_1 &=\Diag(-1, -1, -1, 1),\\
\tmatrix_2 &= \Diag(-1, -1, -\rootunit_4, -\rootunit_4),\\
\tmatrix_3 &= \Diag(-\rootunit_8, -\rootunit_8, -\rootunit_8,-\rootunit_8).
\end{align}
\end{subequations}
Thus, a ququart element in \(\specialsub\) is represented by 
\begin{equation}
X_\permutation
\tmatrix_0^{\ringelement_0}
\tmatrix_1^{\ringelement_1}
\tmatrix_2^{\ringelement_2}
\tmatrix_3^{\ringelement_3},
\end{equation}
where \(\ringelement_0,\ringelement_2\in\Ring_4\), \(\ringelement_1\in\Ring_2\), and \(\ringelement_3\in\Ring_8\).
The computation of the \(\tmatrix_i\) is done using Howell's algorithm~\cite{Howell1986}.

\subsection{Circuit design}
In this subsection we first describe the circuits used for the qutrit and
ququart cases, and then express the sequence fidelity.


The circuit for interleaved benchmarking must result in a expression where the
noises of T and \(\specialsub\) gates are composed.
From this circuit we then estimate the average gate fidelity of the composed
noise.
In practice, the analytical expression for the sequence average must contain the
twirl of the composite channel corresponding to the noise of the T and
\(\specialsub\) gates.

Some requisites of the circuit design are now mentioned.
To ensure that the inversion gate is a member of  \(\specialsub\) we impose one
condition: compute the smallest power of T, name it  \(\powerT\), so that
\(\tmatrix^{\powerT}\in\specialsub\).
Notice \(\powerT\) depends on the dimension: for qutrits and six-level systems,
\(\powerT = 3\);
for even-level systems, \(\powerT = 2\); 
for odd-prime level systems (qupits), \(\powerT = 1\).

We describe some assumptions of our method.
Assume the noise of \(\tmatrix\) is  \(\noiseT\) and the noise for a
member of \(\specialsub\) is  \(\noiseC\).
We further assume the noise of \(\tmatrix\) acts before  \(\tmatrix\)
while the noise of a member \(\groupelement\in\specialsub\) acts after
\(\groupelement\).
We also assume that \(T^{\powerT}\) has noise \(\noiseT\).

\begin{figure}
  \centering
\includegraphics{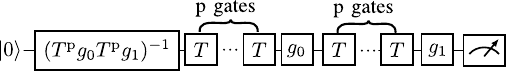}
\caption{
Circuit diagram corresponding to Eq.~\eqref{eq:circuit},
showing an example of the circuits used for our scheme~\cite{MikeAndIke}.
The gates \(\groupelement_i\), for  \(i\in \{0,1\}\), are randomly sampled;
see Sec.~\ref{sec:numerical-results} for a description of a run of our scheme.
We follow the convention time increases to the right~\cite{MikeAndIke}.
This circuit is used to estimate the average gate fidelity of the
\(T\) gate as explained in \S\ref{sec:results}.
}
\label{fig:circuit-construction}
\end{figure}

The circuits required in our scheme are of the form
\begin{equation}\label{eq:circuit}
\composite_i \channel_{\tmatrix^{\powerT}}  \channel_{\groupelement_i},
\end{equation}
for \(\groupelement_i\in\specialsub\).
A schematics is given in Fig.~\ref{fig:circuit-construction}.
Considering noise, we get
\begin{equation}\label{eq:circuit-without-inversion}
\composite_i 
\channel_{\tmatrix^{\powerT}}\noiseT \noiseC  \channel_{\groupelement_i}.
\end{equation}
The inversion gate is
\begin{equation}
\channel_{(\composite_i \tmatrix^{\powerT} \groupelement_i)^{-1}},
\end{equation}
which is a member of the \(\specialsub\) gate set already benchmarked.

We now discuss the resulting operator after averaging over~\(\specialsub\).
Averaging over \(\specialsub\)---using the  elements \(\groupelement_i\)---we obtain
\begin{equation}\label{eq:twirl}
  \noiseC \twirlop^{\circuitdepth},
\end{equation}
with \(\twirlop\) given in Eq.~\eqref{eq:twirl-explicit-form}.

With Eq.~\eqref{eq:twirl} we conclude the construction of the circuits that 
result in the twirl of \(\noiseT\noiseC\),
therefore permitting the computation of the AGF of 
\(\noiseT\noiseC\) in a RB experiment.





\section{Results}\label{sec:results}
In this section,
we write the expressions used to characterise a T gate using our
qudit interleaved benchmarking scheme.
We start by writing the twirl by the Clifford-like gates.
Then the sequence fidelity.
We conclude with the average gate fidelity to characterise a T gate, computed
using Kimmel's approximation~\cite{Kimmel2014}.


The twirl in Eq.~\eqref{eq:twirl} has the following diagonal
form~\cite{amaro2024a}
\begin{equation}\label{eq:twirl-explicit-form}
\twirlop
= 
1 \oplus \eigenvalue_0^{\oplus \qudim-1}\oplus \eigenvalue_+^{\oplus
\qudim^2-\qudim};
\end{equation}
we shortly describe how to estimate the parameters \(\eigenvalue_0\) and
\(\eigenvalue_+\).
By using, in turn, the states \(\ket{\irrepindex}\),
where \(\irrepindex \coloneqq \{0,+\}\) and~\(\ket{+} \coloneqq H\ket0\),
we compute a sequence fidelity,
which it depends only on one of the two parameters in 
Eq.~\eqref{eq:twirl-explicit-form}.

Using the twirl in Eq.~\eqref{eq:twirl-explicit-form}
and observing that the states~\(\ket{\irrepindex}\) lie in the appropriate
invariant subspace~\cite{amaro2024a},
we compute the sequence fidelity.
The sequence fidelity is
\begin{equation}
\Pr(\circuitdepth; \irrepindex)
=
\bbra{\irrepindex}
\noiseC
\twirlop^{\circuitdepth}
\kket{\irrepindex}
=
\spampara + \spamparb \eigenvalue_{\irrepindex}^{\circuitdepth},
\end{equation}
where \(a,b\) are two parameters absorbing state preparation and measurement (SPAM) contributions. These SPAM
contributions are
irrelevant to estimate \(\Fidelity(\channel)\).

The expression for the average gate fidelity of a given \(\channel\)
twirled by \(\specialsub\) is~\cite{NIELSEN2002249}:
\begin{equation}\label{eq:agf}
  \Fidelity_\specialsub(\channel)
  =
  \frac{\qudim(1+(\qudim-1)\eigenvalue_0 +
  (\qudim^2-\qudim)\eigenvalue_+)+\qudim^2}{\qudim^2(\qudim+1)}.
\end{equation}
Therefore, by estimating \(\eigenvalue_\irrepindex\) using the expression for
the sequence fidelity, we compute the average gate fidelity of \(\channel\).

We then can use the expression for the product of fidelities to estimate
the fidelity of the T gate; this expression is one of our key results.
We know the value \(\Fidelity(\channel) = \frac{d
\chientry(\channel)-1}{d+1}\).
Also recall the approximation, for any pair of channels \(\channel\)~and~\(\channel'\),
\(\chientry(\channel\channel')\approx\chientry(\channel)\chientry(\channel')\).
Using this approximation, we estimate \(\chientry(\noiseT)
\approx \chientry(\noiseT
\noiseC)/\chientry(\noiseC)\).
Then we insert the value \(\chientry(\noiseT)\) in \(\Fidelity(\noiseT)\).

\subsection{Numerical results}\label{sec:numerical-results}
In this subsection we illustrate our procedure applied for the qutrit case.
We consider the case the gates in the base gate set have a noise a 
CPTP channel
and the T gate  with fidelities \(0.999\) and \(0.95\), respectively.
In our numerical example we used random channels~\cite{kukulski2021};
this removes any bias our noise could have.

\begin{figure}
  \begin{center}
    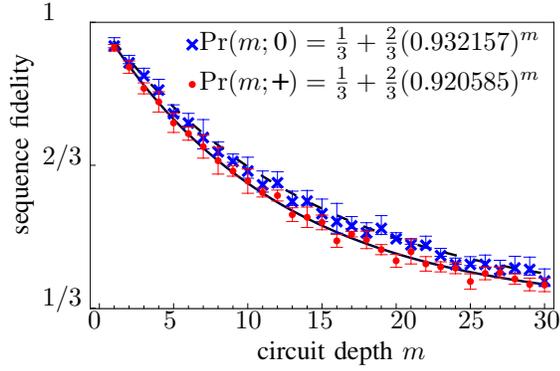
  \end{center}

\caption{This plot presents the fitting procedure applied to numerical results
for a specific configuration in a randomized benchmarking experiment.
The solid line corresponds to the fitting \(\Pr(m;+)\)
and the dashed line to the fitting  \(\Pr(m;0)\).
We
assume prior knowledge of the fidelity of the error for the base gates. The
plot includes two channels, with one channel exhibiting a fidelity, as given in
Eq.~\eqref{eq:simple-formula}, of 
\(0.9996\) and the other \(0.95\).
The fitted model, following the prescribed equation,
incorporates the parameters \(\eigenvalue_\irrepindex\), which are written in
Eq.~\eqref{eq:twirl-explicit-form}.
These
parameters are utilized to estimate the average gate fidelity, as demonstrated
by the numerical example depicted in this figure. The fitting procedure and the
subsequent fidelity estimation are crucial for understanding and optimizing
quantum gate operations within this experimental setup.}
\label{fig:numerical-sequence-fidelity}
\end{figure}

We explain the numerical estimate of one point of 
Fig.~\ref{fig:numerical-sequence-fidelity},
for instance
\(\circuitdepth=2\).
Our explanation illustrates the construction of the circuit in 
Fig.~\ref{fig:circuit-construction} and written in Eq.~\eqref{fig:circuit-construction}.

First the state  \(\ket{0}\) is prepared.
Then, uniformly at random, 
\(m=2\) gates
are sampled from the base gate set: \(g_0,g_1\).
The following sequence of gates is applied to the state \(\ket{0}\) (including
noise).
\begin{equation}
  \noiseC (T^{p}g_0 T^{p}g_1)^{-1}
  T^{p}\noiseT \noiseC g_0
  T^{p}\noiseT \noiseC g_1 \ket{0}.
\end{equation}
Then a measurement with respect to \(\ket{0}\) is done:
\begin{equation}
  \bra{0}
  \noiseC (T^{p}g_0 T^{p}g_1)^{-1}
  T^{p}\noiseT \noiseC g_0
  T^{p}\noiseT \noiseC g_1 \ket{0}.
\end{equation}
We repeated this process \(3\) times to obtain one point in
Fig.~\ref{fig:numerical-sequence-fidelity}.

The next step is to use the values of \(\eigenvalue_\irrepindex\), estimated from the non-linear fit from
Fig.~\ref{fig:numerical-sequence-fidelity}, to compute the average gate
fidelity.
To compute the fidelity, we use Eq.~\eqref{eq:agf}---we obtain \(F = 0.948985\)
and dividing by the fidelity of the already characterised gate set (\(F(\noiseC) =
0.9996\)) we obtain  \(F = 0.949365\). 
Compared to \(F(\noiseT) = 0.95\), the computed fidelity has a percentage error of \(0.1\%\).

\section{Conclusions}
We introduced a scheme to individually characterise qudit non-Clifford
gates.
We provided an explicit circuit design that allows the estimate the average
gate fidelity of the noise composition for the T gate and members of a special
subgroup of the Clifford group.
With the expressions so obtained, a qudit T gate can be individually characterised.
This is a conceptual and practical improvement with respect to previous methods, where the average gate
fidelity was estimated as the average fidelity over the whole gate set.
Our method then allows an appropriate characterisation of a universal qudit gate
set.

\section*{Acknowledgment}

We acknowledge support from Natural Sciences and Engineering Research Council of Canada and Government of Alberta.





\bibliographystyle{IEEEtran}
\bibliography{manuscript}


\end{document}